# Enhancing biomechanical stimulated Brillouin scattering imaging with physics-driven model selection

Roni Shaashoua,[1,*] Tal Levy,[2] Barak Rotblat,[2,3] & Alberto Bilenca[1,4,*]

[1]Biomedical Engineering Department, Ben-Gurion University of the Negev, 1 Ben Gurion Blvd, Be'er-Sheva 84105, Israel

[2]Life Sciences Department, Ben-Gurion University of the Negev, 1 Ben Gurion Blvd, Be'er-Sheva 84105, Israel

[3]The National Institute for Biotechnology in the Negev, Ben-Gurion University of the Negev, 1 Ben Gurion Blvd, Be'er-Sheva 84105, Israel

[4]Ilse Katz Institute for Nanoscale Science and Technology, Ben-Gurion University of the Negev, 1 Ben Gurion Blvd, Be'er-Sheva 84105, Israel

e-mail: *ronishaa@post.bgu.ac.il, bilenca@bgu.ac.il

**Abstract**

Brillouin microscopy is an emerging technique for all-optical biomechanical imaging without the need for physical contact with the sample or for an external mechanical stimulus. However, Brillouin microscopy often retrieves a single, averaged Brillouin frequency shift of all the materials in the sampling volume, introducing significant spectral artifacts in the Brillouin shift images produced. To enable the identification between single- and multi-peak Brillouin signatures in the sample voxels, we developed here a new physics-driven model selection framework based on information theory and an overfit Brillouin water peak threshold. The model selection framework was applied to Brillouin data of NIH/3T3 cells measured by stimulated Brillouin scattering microscopy, facilitating the improved quantification of the Brillouin shift of different regions in the cells, and substantially minimizing spectral artifacts in their Brillouin shift images.

**Introduction**

Brillouin microscopy (BM) is a label-free, all-optical technique for imaging the mechanical properties of biological samples with no physical contact or external mechanical stimulus. BM is based on the phenomenon of Brillouin light scattering, where light is inelastically scattered from spontaneous or stimulated sound waves of frequency Ω and wavevector **q**, which depend on the light wavenumber, scattering angle, and the speed of sound in the medium[1]. BM typically employs a 180° scattering angle, accessing the gigahertz-frequency complex longitudinal modulus, and has found widespread applications in biomechanics, including microscopy and imaging of living cells, organisms, and mammals, flow cytometry, ophthalmologic imaging, and tissue microscopy and endoscopy[2-12].

In BM, it is common to model the Brillouin spectrum measured in a voxel as a single peak line shape, limiting the spectral information extracted—typically the Brillouin frequency shift—to the average of all Brillouin signatures of materials inside the voxel. Although modeling of Brillouin spectra with a multi-peak line shape was used in BM[7,8,11], it was either restricted to discrete positions across the sample or was missing a robust way to select the best-fitting model among the single and multi-Lorentzian line shape models. Importantly, the ability to detect multi-peak Brillouin signatures in voxels of biological cells demands that the signal-to-noise ratio, the imaging contrast, and the spatiospectral resolution all be high. While these requirements are often provided by spontaneous BM at hundreds of milliseconds to seconds[7,8], they are readily offered by stimulated BM at a few tens of milliseconds[11].

Here, we developed a physics-driven model selection strategy using the Akaike information criterion (AIC) and an overfit Brillouin water peak threshold for choosing the simplest but more accurate estimating model, out of the single and multi-Lorentzian line shape models, as a reliable representation of the Brillouin spectrum measured voxelwise in cells. The model selection strategy was applied to Brillouin spectra of living NIH/3T3 cells obtained by stimulated Brillouin scattering (SBS) microscopy, facilitating the improved quantification of the Brillouin shift of different regions in the cell.

**Results and discussion**

Our physics-driven model selection framework utilizes the Akaike information criterion (AIC) which determines the optimum trade-off between model accuracy and model complexity (Materials and methods and Supplementary Note 1). For reducing the risk of overfitting in the model selected, a spectral threshold—related to the linewidth of the Brillouin peak of water—was employed on the Brillouin peaks measured in the cells. It is helpful to understand this threshold by considering the Brillouin data of NIH/3T3 cells (Fig. 1 and Supplementary Fig. 1a,b). The Brillouin spectra were acquired by a stimulated Brillouin scattering (SBS) microscope (Materials and methods and Supplementary Fig. 2) with no apparent photodamage (Supplementary Note 2 and Supplementary Fig. 3). Single and double Brillouin peaks can clearly be identified across the entire cell (orange dots in Fig. 1 and Supplementary Fig. 1b), with the frequency difference between the two peaks of double Lorentzian fits to the Brillouin spectra (hereafter referred as the peak frequency difference) showing minimum values around ~180-220 MHz (Fig. 2a and Supplementary Fig. 1c, blue). This level corresponds approximately to one half the linewidth of the Brillouin water peak (~190 MHz) and roughly matches the 95% percentile of the peak frequency difference of a double Lorentzian line shape overfitted to the Brillouin spectrum of water (207 MHz; Supplementary Note 3 and Supplementary Fig. 4). As biological cells have high water content, we set the threshold for reducing the overfitting to a multi-Lorentzian line shape at 207 MHz. Interestingly, regions with large linewidth $\Gamma_B$ and small peak gain $G_B$ retrieved from the single Lorentzian fits (Fig. 2b,c and Supplementary Fig. 1d,e; red and blue, respectively) overlap well with regions having a large peak frequency difference (Fig. 2a and Supplementary Fig. 1c; red), making wide and low Brillouin peaks appropriate candidates for the double Lorentzian model. It is worth noting that a single Lorentzian line shape may represent either a truly single Brillouin peak or an unresolved Brillouin peak. Further, narrow and high Brillouin peaks may have side peaks, suiting better to the double Lorentzian model (middle panel of Supplementary Fig. 1b).

To demonstrate the model selection scheme in BM, we applied it to the Brillouin spectra measured in the NIH/3T3 cells (Fig. 3a and left panel of Supplementary Fig. 1f). The bar plot shows that the double Lorentzian model was preferred over the single Lorentzian model in most of the cell voxels (>90%). The resulting Brillouin shift $\Omega_B$ image is also presented, and when compared with the $\Omega_B$ image produced using the single Lorentzian model only (Fig. 3b and Supplementary Fig. 1a), the significant spectral artifacts in the latter are evident, as manifested by the markedly downshifted Brillouin frequency shift density (Supplementary Fig. 5). While more complex, higher-order Lorentzian models could be used, we found that the difference in the AIC scores between the double and triple Lorentzian models in the cell is considerably smaller than that between the single and double Lorentzian models (Supplementary Note 4 and Supplementary Fig. 6 and 1h), suggesting that the selection between the single and the double Lorentzian models represents a good balance between model complexity and adequacy to the data. For exploring model selection in different regions of the cell, we evaluated the percentage of the single and the double Lorentzian models preferred in the cytoplasm, nucleoli, and nucleoplasm of the cells measured (Fig. 3c and left panel of Supplementary Fig. 1g). The percentage of the double Lorentzian model is found to be highest in the region showing the widest and lowest Brillouin peaks (cytoplasm in Fig. 3c and nucleoli in the left panel of Supplementary Fig. 1g), likely due to contributions of multiple, relatively balanced, biomechanical constituents with overlapping Brillouin bands.

When further comparing the $\Omega_B$ image produced by the single Lorentzian fits (Fig. 3b and Supplementary Fig. 1a) and that created by the model selection scheme (Fig. 3a and left panel of Supplementary Fig. 1f), it appears that the latter is noisier. To reduce noise with minimal impact on the ability to retrieve from the Brillouin spectrum the Brillouin component of the cell, we used a modified double Lorentzian selection scheme (Materials and methods). As the modified scheme assumes that the Brillouin components of the aqueous buffer around the voxel-under-observation are similar, they can be used to reduce the model complexity in the selection of the preferred model, with a slight reduction in the model accuracy compared to the standard double Lorentzian selection scheme (Supplementary Fig. 6 and 1h). Indeed, the resulting $\Omega_B$ image created by the modified model selection scheme (Fig. 4a and right panel of Supplementary Fig. 1f) appears less noisy than that obtained by the standard model selection scheme (Fig. 3a and left panel of Supplementary Fig. 1f), with the percentage of the single and the double Lorentzian fits selected being comparable to that achieved by the standard model selection scheme (Fig. 4b versus 3c and right versus left panels of Supplementary Fig. 1g). These observations are further supported by the violin plots of the nucleoli, nucleoplasm, and cytoplasm of the cells measured (Supplementary Fig. 5b versus 5c and Supplementary Fig. 1j versus 1k).

**Conclusions**

Grounded on information theory and the overfit Brillouin water peak threshold, the model selection scheme developed offers a useful means for balancing between the goodness-of-fit of single or multi Lorentzian models to Brillouin image data of living cells and the number of parameters in the model. As the model selection scheme involves the identification of multi-peak Brillouin signatures in voxels of a biological sample, the high signal-to-noise ratio, imaging contrast, and spectral resolution of the Brillouin microscope are all essential. Together with the multi-parameter contrast and practical imaging times of SBS microscopy in biological settings, the model selection scheme devised may open new possibilities for an improved, more quantitative analysis of Brillouin imaging data of biological systems.

**Materials and methods**

*Model selection framework*

The model selection framework involves the use of Eq. (4) in Supplementary Note 1 for computing the AIC scores, voxel-by-voxel, of the single and the standard or modified double Lorentzian models in the selection schemes ($P$=1 for the single and the modified double Lorentzian models and $P$=2 for the standard double Lorentzian model). The model that passes the spectral threshold and obtains the lowest AIC score is determined as the preferred model for representing the spectrum measured in the voxel.

While the standard double Lorentzian model is the sum of a constant background and two Lorentzian line shapes with different peak gains, Brillouin frequency shifts and linewidths, the modified double Lorentzian model is a two-step model. The first step includes a traditional double Lorentzian least-squares fit of the Brillouin spectra in a 9×9 sliding spatial window around the voxel-under-observation (i.e., the central voxel of the spatial window). The fit parameters of the Lorentzian line shape that corresponds to the aqueous buffer are subsequently median-filtered over the spatial window. In the second step, the median-filtered fit parameters are used for describing the peak of the aqueous buffer in a second fit of the

Brillouin spectrum of the voxel-under-observation to a double Lorentzian line shape with a constant background.

*Stimulated Brillouin scattering (SBS) microscopy*

The SBS microscope used in this work was described previously in detail[11,13,14] (Supplementary Fig. 2a). Briefly, two counter-propagating 780-nm continuous-wave pump and probe beams are focused with two high numerical aperture objectives (0.7, Olympus) onto a common focal point in the sample, providing ~0.3×0.3×2 μm$^3$ spatial resolution. To measure the stimulated Brillouin gain (SBG) spectrum at the spot, the frequency detuning between the lasers is current-scanned over the desired frequency range and the SBG is measured using a high-frequency lock-in scheme with optical filtering of unwanted stray light reflections by an atomic filter and detection by a low-noise transimpedance receiver. Images of the Brillouin shift, linewidth and peak gain are obtained by raster scanning the sample through the microscope focus.

Shot noise limited Brillouin spectra were recorded at voxels of the sample over a 4-GHz frequency range in 20 ms with SNRs of ~30 dB and an estimated effective spectral resolution of ~100 MHz[11]. These SNR levels correspond to Brillouin shift/linewidth/peak-gain precisions of ~7 MHz/15 MHz/7×10$^{-7}$. The overall power on the sample was ~260 mW and the total imaging time was <7 min for ~19-kilopixel images.

*Cell culture and sample preparation for SBS imaging and the photodamage measurements*

Frozen NIH/3T3 cells (ATCC) were a kind gift from the Mark Schvartzman research group. Cells were maintained using standard tissue culture procedures in a humidified incubator at 37°C with 5% $CO_2$ and atmospheric oxygen. They were grown in a 10 cm cell culture plate in 8 ml Dulbecco's modified Eagle medium (DMEM) with 4.5 g/ml glucose (Biological Industries). The DMEM was supplemented with 10% Fatale Bovine Serum (FBS) (Biological Industries) and 1% Antibiotic-Antimycotic (Anti-Anti). Cells were passaged by taking the media out and washing with 2 ml PBS (Biological Industries). Next, 1.5 ml of 0.25% trypsin and 0.002% EDTA solution (Biological Industries) were added to them. The cells were then incubated at 37°C with 5% $CO_2$ and atmospheric oxygen for 6 minutes until they were detached from the plate. 3.5 ml of DMEM was added to the detached cells and 0.5 ml of it was added into a new 10 cm plate with 8 ml DMEM. Cells were seeded at the amount of ~15,000 cells on top of a 7-mm diameter collagen-coated polyacrylamide gel layer with a shear modulus of 31 kPa attached to a 25-mm diameter glass coverslip[2]. The cells were kept in the incubator for 24 hours in phenol red free DMEM medium (Biological Industries). Immediately prior to SBS imaging and the photodamage measurements, cells were sandwiched between two glass coverslips with an aqueous buffer in the sample holder (Supplementary Fig. 2b).

**References**

1. Boyd, RW., *Nonlinear Optics 3rd ed. (Academic Press, New York)*, pp. 402-412 (2008).

2. Scarcelli, G., Polacheck, WJ., Nia, HT., Patel, K., Grodzinsky, AJ., Kamm, RD. & Yun, SH. Noncontact three-dimensional mapping of intracellular hydromechanical properties by Brillouin microscopy. *Nat Methods* **12**, 1132-1134 (2015).

3. Antonacci, G., Pedrigi, RM., Kondiboyina, A., Mehta, VV., de Silva, R., Paterson, C., Krams, R. & Török, P. Quantification of plaque stiffness by Brillouin microscopy in experimental thin cap fibroatheroma. *J R Soc Interface* **12**, 20150843 (2015).

4. Meng, Z., Traverso, AJ., Ballmann, CW., Troyanova-Wood, MA. & Yakovlev, VV. Seeing cells in a new light: a renaissance of Brillouin spectroscopy. *Adv Opt Photon* **8**, 300-327 (2016).

5. Elsayad, K., Werner, S., Gallemi, M., Kong, J., Sanchez Guajardo, ER., Zhang, L., Jaillais, Y., Greb, T. & Khadir, Y. Mapping the subcellular mechanical properties of live cells in tissues with fluorescence emission–Brillouin imaging. *Sci Signal* **9**, rs5 (2016).

6. Schlüßler, R., Möllmert, S., Abuhattum, S., Cojoc, G., Müller, P., Kim, K., Möckel, C., Zimmermann, C., Czarske, J. & Guck, J. Mechanical mapping of spinal cord growth and repair in living zebrafish larvae by Brillouin imaging. *Biophys J* **115**, 911-923 (2018).

7. Mattana, S., Mattarelli, M., Urbanelli, L., Sagini, K., Emiliani, C., Dalla Serra, M., Fioretto, D. & Caponi, S. Non-contact mechanical and chemical analysis of single living cells by microspectroscopic techniques. *Light Sci Appl* **7**, 17139 (2018).

8. Bevilacqua, C., Sánchez-Iranzo, H., Richter, D., Diz-Muñoz, A. & Prevedel, R. Imaging mechanical properties of sub-micron ECM in live zebrafish using Brillouin microscopy. *Biomed Opt Express* **10**, 1420-1431 (2019).

9. Margueritat, J., Virgone-Carlotta, A., Monnier, S., Delanoë-Ayari, H., Mertani, HC., Berthelot, A., Martinet, Q., Dagany, X., Rivière, C., Rieu, JP. & Dehoux,T. High-frequency mechanical properties of tumors measured by Brillouin light scattering. *Phys Rev Lett* **122**, 018101 (2019).

10. Bailey, M., Alunni-Cardinali, M., Correa, N., Caponi, S., Holsgrove, T., Barr, H., Stone, N., Winlove, CP., Fioretto, D. & Palombo, F. Viscoelastic properties of biopolymer hydrogels determined by Brillouin spectroscopy: A probe of tissue micromechanics. *Sci Adv* **6**, eabc1937 (2020).

11. Remer, I., Shaashoua, R., Shemesh, N., Ben-Zvi, A. & Bilenca, A. High-sensitivity and high-specificity biomechanical imaging by stimulated Brillouin scattering microscopy. *Nat Methods* **17**, 913-916 (2020).

12. Poon, C., Chou, J., Cortie, M. & Kabakova, I. Brillouin imaging for studies of micromechanics in biology and biomedicine: from current state-of-the-art to future clinical translation. *J Phys Photonics* **3,** 012002 (2021).

13. Remer, I. & Bilenca, A. High-speed stimulated Brillouin scattering spectroscopy at 780 nm. *APL Photonics* **1**, 061301 (2016).

14. Remer, I., Cohen, L. & Bilenca, A. High-speed continuous-wave stimulated Brillouin scattering spectrometer for material analysis. *J Vis Exp* **127**, 55527, (2017).

**Acknowledgments**

A.B. acknowledges the support of the Israel Science Foundation (grant no. 1173/17). We would like to thank Prof. Giuliano Scarcelli and Miloš Nikolić for sharing their protocol and advice on the preparation of the cell samples.

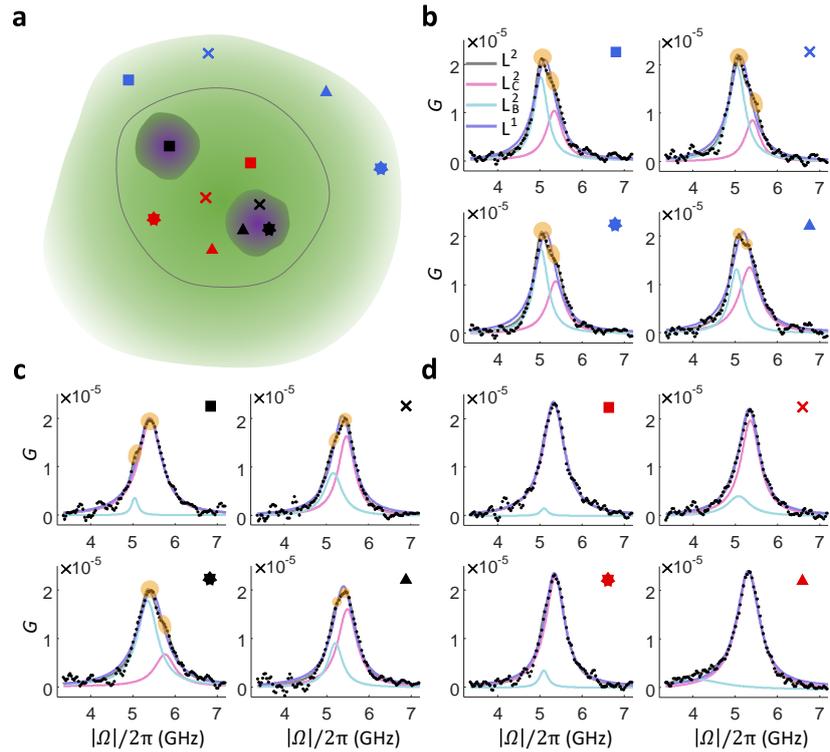

**Fig. 1 | Representative stimulated Brillouin gain (SBG) spectra of an NIH/3T3 cell. a,** Schematic of the multiple sites sampled across the cell indicated by square, star, triangle, and cross symbols. Blue, cytoplasm; black, nucleoli; red, nucleoplasm. **b-d,** Representative SBG spectra of the cytoplasm (**b**), nucleoli (**c**), and nucleoplasm (**d**). Multiple peaks are marked with orange dots. Measured spectra are shown in black dots, and single and double Lorentzian models in purple ($L^1$) and gray ($L^2$) solid lines, with the Brillouin component of the cell in magenta ($L_C^2$) and that of the aqueous buffer in cyan ($L_B^2$). All spectra were measured over 4 GHz within 20 ms. The overall power on the sample was ~260 mW.

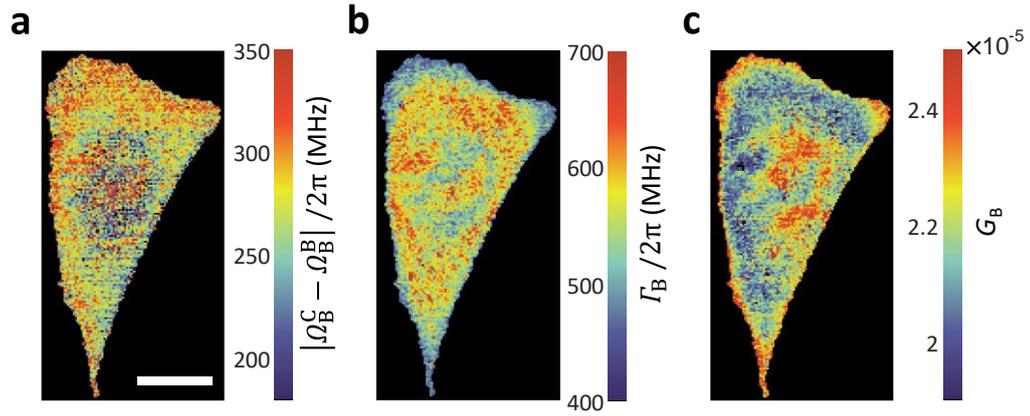

**Fig. 2 | Brillouin peak structure in an NIH/3T3 cell. a,** Image of the peak frequency difference. **b, c,** Images of the Brillouin linewidth $\Gamma_B$ (**b**) and peak gain $G_B$ (**c**) obtained using single Lorentzian fits. All SBG spectra were measured over 4 GHz within 20 ms. The overall power on the sample was ~260 mW. Image size, ~100×190 pixels; pixel size, 0.25×0.25 μm$^2$; Scale bar, 10 μm.

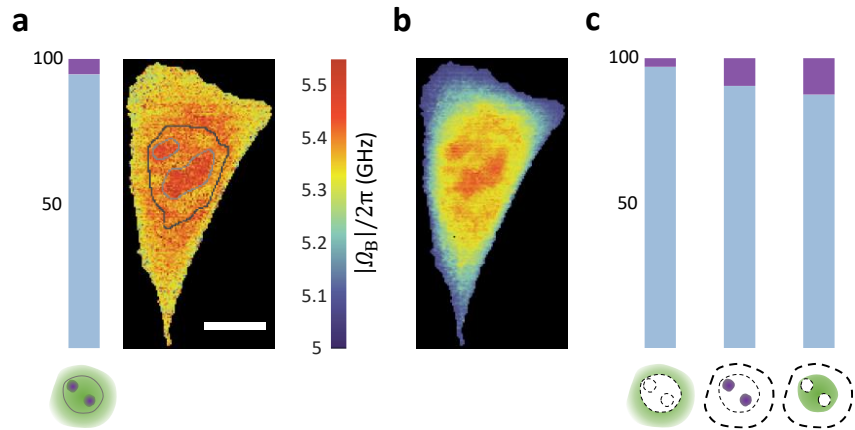

**Fig. 3 | Physics-driven model selection in Brillouin microscopy of cells. a,** Bar plot of the percentage of the single (purple) and double (light blue) Lorentzian models selected as the preferred model in the cell voxels along with the resulting Brillouin shift image. **b,** Brillouin shift image produced by single Lorentzian fits. **c,** Bar plots of the percentage of the single (purple) and the double (light blue) Lorentzian models selected as the preferred model in the cytoplasm, nucleoli, and nucleoplasm (left to right). Imaging and image parameters are as in Fig. 2.

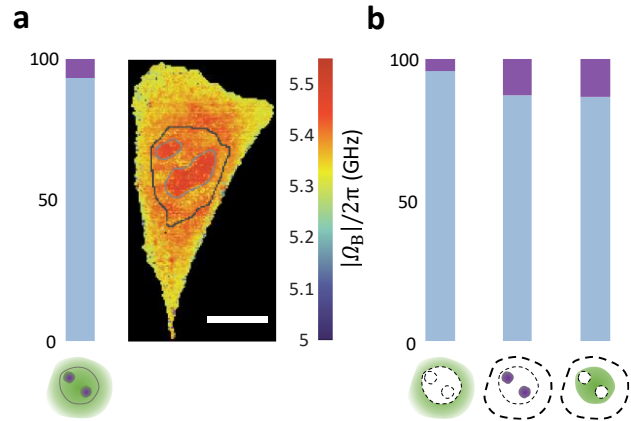

**Fig. 4 | Physics-driven modified model selection in Brillouin microscopy of cells. a,** Bar plot of the percentage of the single (purple) and modified double (light blue) Lorentzian models selected as the preferred model in the cell voxels along with the resulting Brillouin shift image. **b,** Bar plots of the percentage of the single (purple) and the modified double (light blue) Lorentzian models selected as the preferred model in the cytoplasm, nucleoli, and nucleoplasm (left to right). Imaging and image parameters are as in Fig. 2.

Supplementary Information

# Enhancing biomechanical stimulated Brillouin scattering microscopy with physics-driven model selection

Roni Shaashoua,[1,*] Tal Levy,[2] Barak Rotblat,[2,3] & Alberto Bilenca[1,4,*]

[1]Biomedical Engineering Department, Ben-Gurion University of the Negev, 1 Ben Gurion Blvd, Be'er-Sheva 84105, Israel

[2]Life Sciences Department, Ben-Gurion University of the Negev, 1 Ben Gurion Blvd, Be'er-Sheva 84105, Israel

[3]The National Institute for Biotechnology in the Negev, Ben-Gurion University of the Negev, 1 Ben Gurion Blvd, Be'er-Sheva 84105, Israel

[4]Ilse Katz Institute for Nanoscale Science and Technology, Ben-Gurion University of the Negev, 1 Ben Gurion Blvd, Be'er-Sheva 84105, Israel

e-mail: *ronishaa@post.bgu.ac.il, bilenca@bgu.ac.il

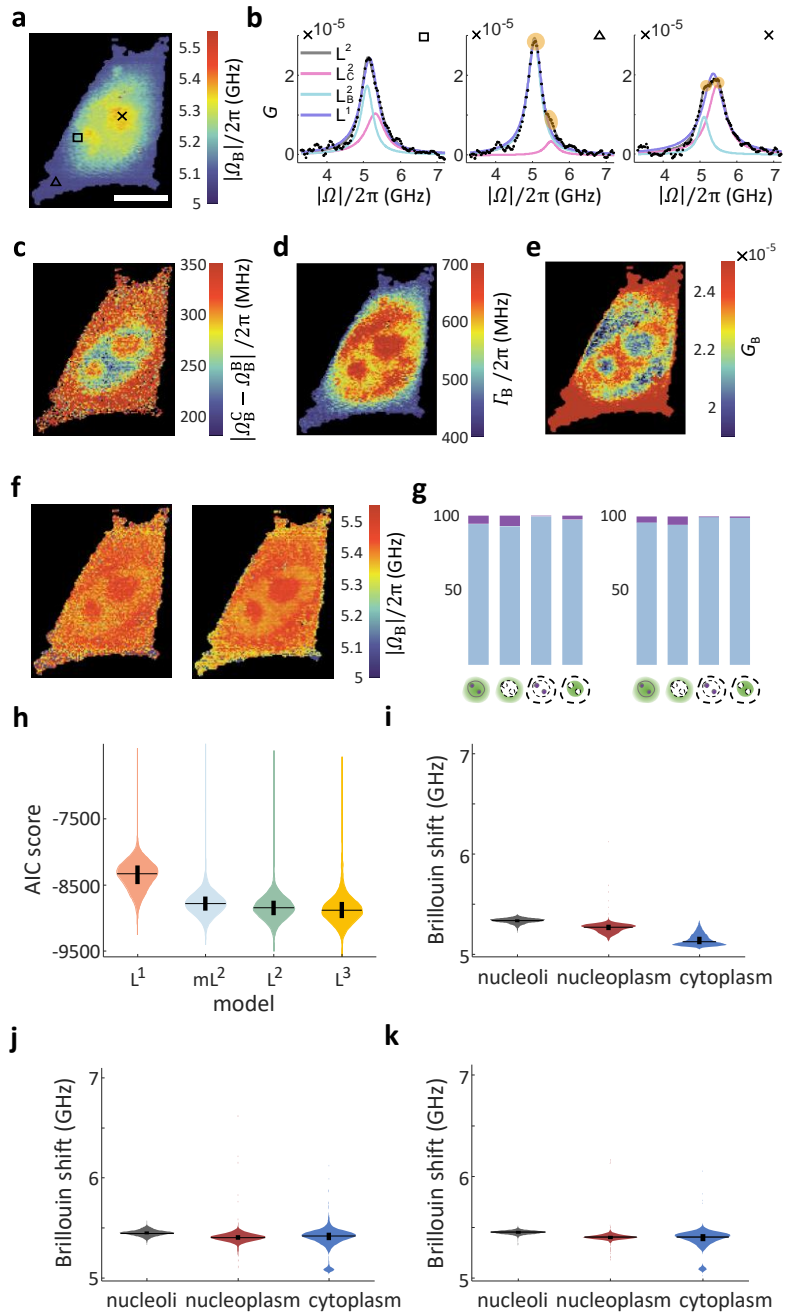

**Supplementary Fig. 1 | Physics-driven model selection in an NIH/3T3 cell. a,** Brillouin shift image produced by single Lorentzian fits. **b,** Representative SBG spectra at different locations across the cell indicated by square, triangle, and cross in **a**. Measured spectra are shown in black dots, and single and double Lorentzian models in purple ($L^1$) and gray ($L^2$) solid lines, with the Brillouin component of the cell in magenta ($L_C^2$) and that of the aqueous buffer in cyan ($L_B^2$). **c,** Peak frequency difference. **d, e,** Brillouin linewidth $\Gamma_B$ (**d**) and peak gain $G_B$ (**e**) obtained using single Lorentzian fits. **f, g,** Brillouin shift images (**f**) and percentage bar plots (**g**) using the standard (left) and the modified

(right) double Lorentzian selection schemes. Single and double (or modified double) Lorentzian models selected as the preferred model across the entire cell and in the cell cytoplasm, nucleoli, and nucleoplasm (left to right) are shown in purple and light blue, respectively. **h,** Violin plots of the AIC scores of the single, double, modified double, and triple Lorentzian models ($L^1$, $L^2$, $mL^2$, $L^3$). The AIC scores of the four models were statistically different from each other. **i-k,** Violin plots of the Brillouin shifts of the cell nucleoli, nucleoplasm, and cytoplasm for the single Lorentzian model (**i**) and the standard and modified double Lorentzian selection schemes (**j, k**). Using ANOVA with multiple comparisons, the Brillouin shifts of the three regions were found to be statistically different in **i, k**. In **j**, the *p*-value between the nucleoplasm and cytoplasm regions was 0.0514, whereas the other region combinations were statistically different. All the data was measured over 4 GHz within 20 ms. The overall power on the sample was ~260 mW. Image size, ~100×130 pixels; pixel size, 0.25×0.25 $\mu m^2$; Scale bar, 10 μm.

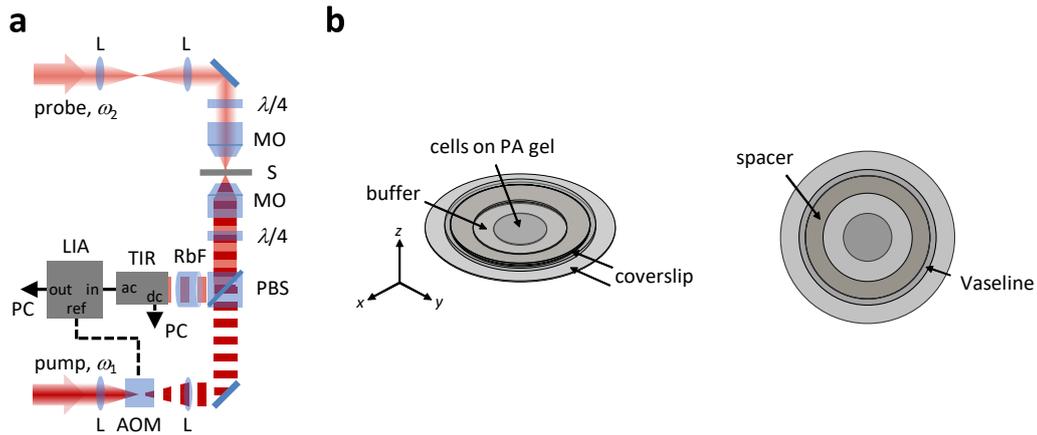

**Supplementary Fig. 2 | SBS microscope and sample holder. a,** For stimulated Brillouin gain (SBG) measurements, the intensity of the pump beam is modulated at 1.1 MHz by an acousto-optic modulator (AOM) and the resulting intensity modulation of the probe beam owing to SBS in the sample (S) is detected at the same frequency with a lock-in amplifier (LIA) following optical filtering by an 85-Rb filter (RbF) and detection using a transimpedance receiver (TIR). For brightfield imaging, the microscope was supplemented with an additional white light source (not shown). L, lens; MO, microscope objective; PBS, polarizing beam splitter; PC, personal computer. **b,** Isometric and top views of the sample holder comprising two 0.15-mm thick glass coverslips of 25 mm and 18 mm in diameter spaced at 0.36-mm thickness using adhesive spacers. Cells were grown on a 0.12-mm thick collagen-coated polyacrylamide gel layer, which was polymerized on the bottom coverslip. The residual space between the glasses was filled with the aqueous buffer. Vaseline ointment is used to ensure complete sealing of the sample.

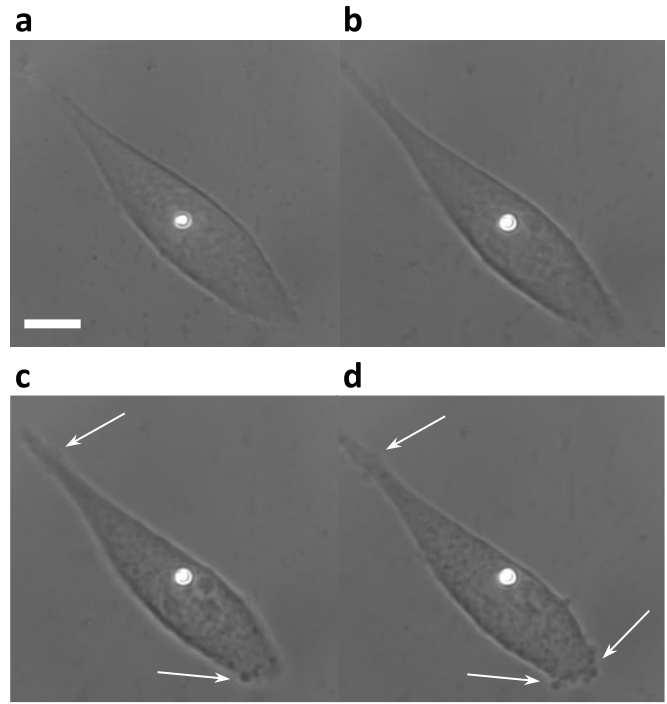

**Supplementary Fig. 3 | Representative brightfield images of an NIH/3T3 cell under focused illumination in the SBS microscope at different times from the beginning of irradiation.** 0 min (**a**), 17 min (**b**), 34 min (**c**), 39 min (**d**). The illumination focusing power and intensity are the same as those used for SBS imaging of the cells. The arrows point on locations where blebs are apparent, indicating some level of cell damage. Scale bar, 10 μm.

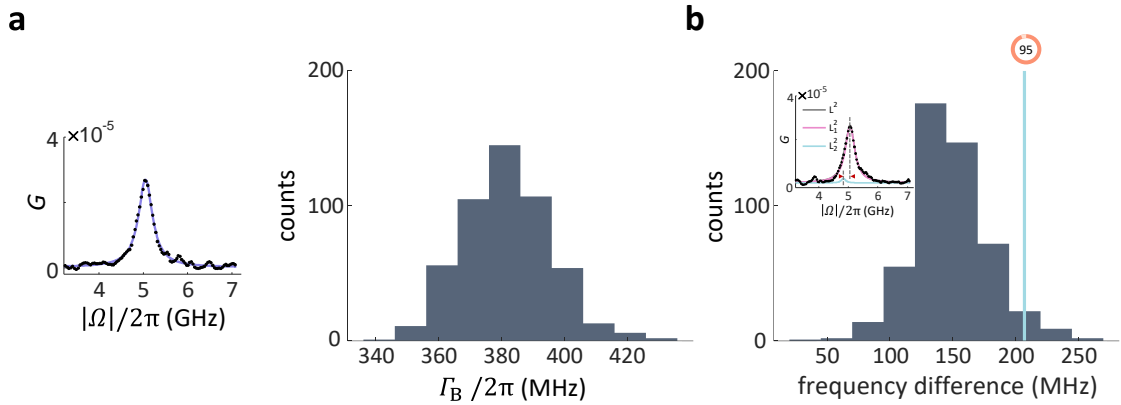

**Supplementary Fig. 4 │ Brillouin water peak measurements. a,** Representative measurement of the Brillouin spectrum of water (black dots) and the single Lorentzian fit (purple solid line) along with the histogram of the linewidth (in full width at half-maximum). **b,** Histogram of the peak frequency difference and the 95th percentile of the frequency difference (vertical line). The inset shows the double Lorentzian overfit to the Brillouin water peak ($L^2$, gray solid line) with the two Lorentzian components ($L_1^2$, magenta; $L_2^2$, cyan). Arrowheads indicate the peak frequency difference.

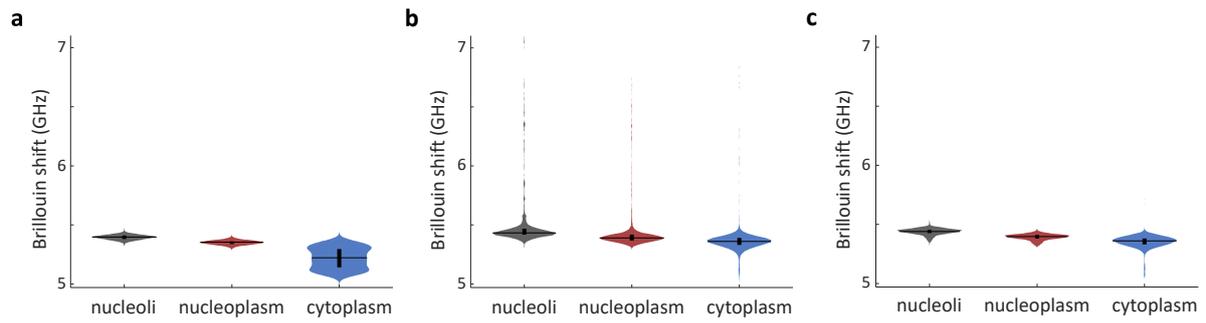

**Supplementary Fig. 5 | Violin plots of the Brillouin shift of the nucleoli, nucleoplasm, and cytoplasm of the cell presented in the main text for the different models and model selection schemes.** Single Lorentzian model (**a**), double Lorentzian model selection (**b**), modified double Lorentzian model selection (**c**). The black horizontal and vertical lines denote the median and interquartile range, respectively. Using ANOVA with multiple comparisons, the Brillouin shifts of the three regions were found to be statistically different in **a-c**.

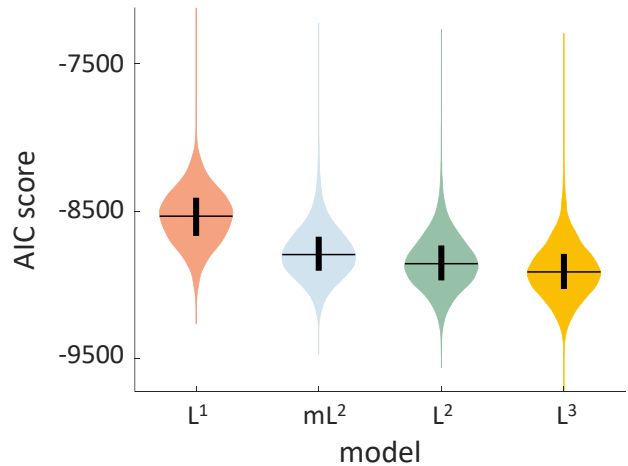

**Supplementary Fig. 6 │ Violin plots of the AIC scores of the single, double, modified double, and triple Lorentzian models calculated for the cell presented in the main text.** The single Lorentzian model $L^1$ has 4 unknown model parameters (frequency shift, linewidth, and peak gain of the spectral Lorentzian line shape and a background constant). The double Lorentzian model $L^2$ has 7 unknown model parameters (frequency shift, linewidth, and peak gain of each of the two Lorentzian components and a background constant). The modified double Lorentzian model $mL^2$ has 3 known model parameters (median frequency shift, linewidth, and peak gain of the Lorentzian component of the aqueous buffer around the central voxel-under-observation in a 9×9 sliding spatial window) and 4 unknown model parameters (frequency shift, linewidth, and peak gain of the Lorentzian component of the cell and a background constant). The triple Lorentzian model $L^3$ comprises 10 unknown parameters (frequency shift, linewidth, and peak gain of each of the three Lorentzian components and a background constant). The black horizontal and vertical lines denote the median and interquartile range, respectively. The AIC scores of the four models were statistically different from each other.

**Supplementary Note 1: Akaike information criterion (AIC) in Brillouin microscopy**

AIC is one of the most widely used approaches for selecting the best-fitting model from several candidate models given a particular data set. The AIC score reads as[1]

$$-2\ln\left(\mathrm{L}\left(\hat{\boldsymbol{\theta}}_{\mathrm{MLE}}\mid\boldsymbol{y}\right)\right)+2K, \tag{1}$$

where $\hat{\boldsymbol{\theta}}_{\mathrm{MLE}}$ is the maximum likelihood estimator (MLE) of the parameter vector, $\ln\left(\mathrm{L}\left(\hat{\boldsymbol{\theta}}_{\mathrm{MLE}}\mid\boldsymbol{y}\right)\right)$ is the log-likelihood of $\hat{\boldsymbol{\theta}}_{\mathrm{MLE}}$ given the measurement vector $\boldsymbol{y}$, and $K$ is the total number of estimated parameters. The first term in Eq. (1) is a measure of the fit accuracy of the candidate model and increases (in absolute value) with the increase in the likelihood of the model, whereas the second term penalizes more complex candidate models. Therefore, the candidate model which has the lowest AIC value is the preferred model for representing the measured spectrum.

To calculate the AIC score of the $P$ Lorentzian model, we first assume that the statistical model of the spectrum is $\boldsymbol{y} = B^P + \boldsymbol{L}\boldsymbol{G}_{\mathbf{B}}^P + \boldsymbol{n}$ (Supplementary Ref.[2]), where $B^P$ is the measurement background, $\boldsymbol{L}_{N\times P}$ is a matrix whose columns represent $P$ Lorentzian line shapes sampled at frequencies $\{\omega_j\}_{j=1}^{N}$ with Brillouin frequency shifts $\{\Omega_{\mathrm{B},p}^P\}_{p=1}^{P}$ and linewidths $\{\Gamma_{\mathrm{B},p}^P\}_{p=1}^{P}$, $\boldsymbol{G}_{\mathbf{B}_{P\times 1}}^P$ is a vector describing the Brillouin peak-gain of the $p$-th Lorentzian line shape, and $\boldsymbol{n}_{N\times 1}$ is the Gaussian noise vector with zero mean and $\sigma^2 \boldsymbol{I}_N$ covariance matrix. Then, the term $\ln\left(\mathrm{L}\left(\hat{\boldsymbol{\theta}}_{\mathrm{MLE}}\mid\boldsymbol{y}\right)\right)$ can readily be expressed as

$$-\frac{N}{2}\ln(2\pi) - N\ln\sigma - \frac{\left\|\boldsymbol{y} - \left(B^P + \boldsymbol{L}\boldsymbol{G}_{\mathbf{B}}^P\right)\right\|^2}{2\sigma^2}. \tag{2}$$

To obtain the first term in Eq. (1), the MLEs of the deterministic model parameters $\theta_{\text{model}}^{P}=\left[B^{P},\ G_{\mathbf{B}}^{P},\ \Omega_{\mathbf{B}}^{P},\ \Gamma_{\mathbf{B}}^{P}\right]_{(3P+1)\times 1}$ are obtained as the solution of the least squares problem

$$\hat{\theta}_{\text{model MLE}}^{P} = \left[\hat{B}_{\text{MLE}}^{P},\ \hat{G}_{\mathbf{B}_{\text{MLE}}}^{P},\ \hat{\Omega}_{\mathbf{B}_{\text{MLE}}}^{P},\ \hat{\Gamma}_{\mathbf{B}_{\text{MLE}}}^{P}\right] = \arg\min_{\left[B^{P},\ G_{\mathbf{B}}^{P},\ \Omega_{\mathbf{B}}^{P},\ \Gamma_{\mathbf{B}}^{P}\right]} \left\| y - \left(B^{P} + L\left(\Omega_{\mathbf{B}}^{P},\ \Gamma_{\mathbf{B}}^{P}\right)G_{\mathbf{B}}^{P}\right)\right\|^{2}.$$

The MLE of the statistical model parameter $\theta_{\text{statistical}}=\sigma$ is calculated by maximizing Eq. (2) with respect to $\sigma$ given the MLEs of the deterministic model parameters $\hat{\theta}_{\text{model}}$, yielding

$$\hat{\sigma}_{\text{MLE}}^{2} = \frac{1}{N}\left\| y - \left(\hat{B}_{\text{MLE}}^{P} + L\left(\hat{\Omega}_{\mathbf{B}_{\text{MLE}}}^{P},\ \hat{\Gamma}_{\mathbf{B}_{\text{MLE}}}^{P}\right)\hat{G}_{\mathbf{B}_{\text{MLE}}}^{P}\right)\right\|^{2}. \quad (3)$$

By substituting Eq. (2) and $\hat{\theta}_{\text{MLE}} = [\hat{\theta}_{\text{model MLE}}^{P},\ \hat{\sigma}_{\text{MLE}}^{2}]$ in Eq. (1) along with the total number of estimated parameters $K=3P+2$, the AIC score of the $P$ Lorentzian model is given by

$$N\ln\left(\frac{\left\| y - \left(\hat{B}_{\text{MLE}}^{P} + L\left(\hat{\Omega}_{\mathbf{B}_{\text{MLE}}}^{P},\ \hat{\Gamma}_{\mathbf{B}_{\text{MLE}}}^{P}\right)\hat{G}_{\mathbf{B}_{\text{MLE}}}^{P}\right)\right\|^{2}}{N}\right) + 2(3P+2). \quad (4)$$

Note that Eq. (4) omits a constant term which is identical for all the models.

**Supplementary Note 2: Photodamage measurements in NIH/3T3 cells**

Photodamage measurements were conducted by examining the time it takes for the NIH/3T3 cells to start blebbing—a commonly used and reliable morphological sign for cellular photodamage[3]—under the same experimental conditions used for SBS imaging, but without sample scanning. We brightfield imaged nine cells at a frame rate of 0.2 fps using the microscope (Supplementary Fig. 2a) and found that blebs started to appear $30 \pm 8$ min after the beginning of the focused illumination (Supplementary Fig. 3). This result is in good agreement with similar experiments reported in Supplementary Ref.[3].

**Supplementary Note 3: Brillouin water peak measurements**

To measure the Brillouin water peak, 500 SBG spectra of Milli-Q water were acquired at room temperature under the same experimental conditions as used for the cells (Supplementary Fig. 4a). The full-width and half-maximum of the Brillouin water peak extracted from these measurements is $382 \pm 14$ MHz. The measured spectra were subsequently fitted to a double Lorentzian line shape and the histogram of the frequency difference between the two overfitted peaks was computed (Supplementary Fig. 4b). From this histogram, we evaluated the 95% percentile of the peak frequency difference to be 207 MHz.

**Supplementary Note 4: AIC scores of the single and multi Lorentzian models**

The violin plots of the AIC scores of the single, double, modified double, and triple Lorentzian models were computed for the cell shown in the main text (Supplementary Fig. 6 and 1h). Using ANOVA with multiple comparisons, all models were found to be statistically different from each other. Nevertheless, we can observe that the single Lorentzian model is the most distant from the other multi Lorentzian models—which are much closer to each other (in terms of the mean AIC score)—indicating that the selection between the single Lorentzian model and the standard or modified double Lorentzian models is a good balance between model complexity and adequacy to the data.

**Supplementary References**

1. Burnham, KP. & Anderson, DR. *Model Selection and Multimodel Inference: A Practical Information-Theoretic Approach 2nd ed. (Springer, New York)*, pp. 60-64 (2002).

2. Remer, I. & Bilenca, A. Background-free Brillouin spectroscopy in scattering media at 780 nm via stimulated Brillouin scattering. *Opt Lett* **41**, 926-929 (2016).

3. Nikolić, M. & Scarcelli, G. Long-term Brillouin imaging of live cells with reduced absorption-mediated damage at 660nm wavelength. *Biomed Opt Express* **10**, 1567-1580 (2019).